# Structural phase transitions in $Pr_{1-x}La_xAlO_3$: Heat capacity and x-ray scattering studies

Makoto Tachibana,[1,2] Katharina Fritsch,[1] and Bruce D. Gaulin[1,3,4]

[1] Department of Physics and Astronomy, McMaster University, Hamilton, Ontario, L8S 4M1, Canada

[2] National Institute for Materials Science, 1-1 Namiki, Tsukuba, Ibaraki, 305-0044, Japan

[3] Brockhouse Institute for Materials Research, McMaster University, Hamilton, Ontario, L8S 4M1, Canada

[4] Canadian Institute for Advanced Research, 180 Dundas Street West, Toronto, Ontario, M5G 1Z8, Canada

**Abstract**

Structural phase transitions in $Pr_{1-x}La_xAlO_3$ ($0 \leq x \leq 1$) single crystals have been studied through heat capacity and high-resolution x-ray scattering measurements. For $PrAlO_3$, the heat capacity shows a sharp first-order peak at the rhombohedral to orthorhombic transition, while a classic mean-field anomaly is observed at the orthorhombic to monoclinic transition at lower temperatures. The transition temperature and the heat capacity anomaly of the two transitions diminish with increasing $x$, and only a single rhombohedral to monoclinic transition is observed for $x = 0.8$. Although this latter transition is required by group theory to be first order, no such evidence is found in the heat capacity and x-ray scattering measurements. Instead, the results are consistent with mean-field criticality, with a small distribution of transition temperature originating from weak disorder.

# I. INTRODUCTION

There has been continuing interest in perovskite $PrAlO_3$ over the last 40 years, as this system shows an unusual sequence of structural phase transitions. At ~1800 K, $PrAlO_3$ undergoes a second-order cubic (space group $Pm-3m$) to rhombohedral ($R-3c$) transition on cooling.[1] This structural modification, which was first observed in $LaAlO_3$ at ~800 K,[1] represents a classic zone-boundary soft mode transition where the $AlO_6$ octahedra tilt in a staggered manner along the cubic <111> direction.[1,2] While $LaAlO_3$ retains this structure down to the lowest temperature, $PrAlO_3$ exhibits an additional set of transitions[3,4]: At 210 K, the axis of the tilt shifts to the cubic <101> direction, resulting in a first-order transition to an orthorhombic *Imma* structure. This is followed by a second-order transition to a monoclinic $C2/m$ structure at 151 K, below which the tilt axis continuously approaches (but never quite reaches) the cubic <001> direction. In 1973, Harley and co-workers[5] pointed out that these two transitions can be explained in the framework of cooperative Jahn-Teller transition,[6] which set off an intense effort to understand the coupling between the $Pr^{3+}$ electronic energy levels and the anharmonic phonon instabilities.[7–13] As a result, the second-order transition at 151 K became one of the best-understood structural phase transitions — its order parameter shows classical mean-field behavior, and this is concomitantly followed by a splitting of the $Pr^{3+}$ crystal field levels (from fluorescence,[5] Raman scattering,[5,10,12,13] or optical absorption[3,11,13] measurements), by internal atomic displacements (from electron spin resonance[8,11]), and by a monoclinic strain (from elastic neutron scattering[9]). This remarkable behavior has since been discussed in various review articles, including the monograph by White and Geballe.[14]

More recently, $PrAlO_3$ was reinvestigated[15–17] as part of a broader group-theoretical approach[18–21] to understand octahedral tilting and Jahn-Teller distortions in perovskites. Using neutron powder diffraction,

Carpenter and co-workers[16] examined the coupling between the two structural instabilities in detail, and provided a formal strain analysis of the structural phase transitions. As a natural extension to these developments, the structural phase transitions in the solid-solution system of $Pr_{1-x}La_xAlO_3$ were also reexamined.[22–25] At high temperatures, the $Pm–3m \leftrightarrow R–3c$ transition extends across the entire series, with the transition temperature decreasing monotonically with increasing $x$.[22] Below room temperature, the phase diagram (Fig. 1) shows interesting evolution of the $R–3c \leftrightarrow Imma$ and $Imma \leftrightarrow C2/m$ transitions[13,23,24]: The temperature interval between the two transitions diminishes with increasing $x$, and the intermediate $Imma$ phase disappears at $x \sim 0.75$. For larger $x$, the $R–3c$ structure transforms directly to the $C2/m$ structure,[13,23] providing a rare example of $R–3c \leftrightarrow C2/m$ transition in perovskites. Moreover, the $C2/m$ phase persists up to the very high doping of $x \sim 0.95$, which seems to question the role of the Jahn-Teller effect in stabilizing this monoclinic structure.[25,26] As previous studies on the high doping regime[13,23] were mostly limited to determining the transition temperatures, further experimental studies are needed to understand the unusual transitions in $Pr_{1-x}La_xAlO_3$.

In this paper, we report the results of heat capacity and high-resolution x-ray scattering measurements on single crystals of $Pr_{1-x}La_xAlO_3$. While a wide variety of experiments on $PrAlO_3$ has been reported over the years, heat capacity data is still lacking for this compound. Our heat capacity measurements on $Pr_{1-x}La_xAlO_3$ ($x = 0, 0.2, 0.4, 0.7, 0.8,$ and $1$) show interesting evolution as a function of composition, especially for large $x$ where the two transitions merge into a single transition. To track the order parameter of the $R–3c \leftrightarrow C2/m$ transition found in $x = 0.8$, the temperature dependence of the monoclinic twinning angle was studied using x-ray scattering measurements. Although this transition is required by symmetry requirements to be first order,[18,19] our calorimetric and structural data show no evidence of hysteresis, latent heat, or discontinuity within the resolution of our measurements.

## II. EXPERIMENT

For this study, single crystals of $Pr_{1-x}La_xAlO_3$ ($x$ = 0, 0.2, 0.4, 0.7, and 0.8) were grown by the flux method.[27] We found that best crystals are obtained when appropriate mixtures[27] of $PbO$, $PbO_2$, $PbF_2$, $B_2O_3$, and $MoO_3$ are used as the flux. Notably, the addition of $MoO_3$ improved the growth of pseudo-cubic {100}-type faces, resulting in rectangular crystals with sides up to 3 mm in extent for each composition. The powder x-ray diffraction patterns of the crushed crystals showed no sign of any impurity phase, and inductively coupled plasma analysis confirmed that the Pr:La ratio agrees within ±2% with the starting compositions. For $LaAlO_3$, a single crystal grown by the Czochralski method was obtained from a commercial supplier. Heat capacity measurements between 2 and 300 K were performed by the relaxation method using a Quantum Design physical properties measurement system. The single crystal x-ray scattering measurements were performed using Cu $K\alpha_1$ radiation, obtained with an 18 kW rotating anode generator and a perfect single crystal Ge (111) monochromator. The sample was mounted on the cold finger of a closed cycle refrigerator and aligned within a four-circle diffractometer. The temperature dependence of the monoclinic twinning angle for $x$ = 0.8 was measured on cooling, with the temperature stability of 0.01 K.

## III. RESULTS AND DISCUSSION

### A. Heat capacity measurements

The heat capacity ($C_p$) data for $PrAlO_3$ and $LaAlO_3$, obtained through measurements on heating, are shown in Fig. 2. For $PrAlO_3$, we find a sharp peak at 212 K and a mean-field peak at 151 K. The sharp peak, from the $R\text{--}3c \leftrightarrow Imma$ transition, was replaced by a much broader anomaly when the measurements were performed on cooling. Also, due to the presence of a latent heat,[28,29] the temperature relaxation after the application of heat pulse was not strictly exponential near the transition. These observations are all consistent with the transition being strongly first order in character. Because the $C_p$ values were obtained by fitting the relaxation curve to an exponential function, the present measurements may underestimate the height of the first-order peak.[28]

The mean-field-type $C_p$ anomaly for the $Imma \leftrightarrow C2/m$ transition agrees with the results of previous studies, where mean-field behavior was reported for the temperature dependence of the order parameter.[8–11] As expected for a second-order transition, there was no sign of latent heat or hysteresis across this anomaly. When a smooth background, shown as a dashed line, is subtracted from the data, we obtain the excess $\Delta C_p$ shown in the inset. Here, the mean-field behavior is immediately recognized as an exponential rise on the low-temperature side, and a sharp drop above the transition. (The transition temperature of 151 K corresponds to the midpoint of the sharp drop.) The $\Delta C_p$ jump at the transition is 17.7 J K$^{-1}$mol$^{-1}$. This mean-field behavior is usually ascribed to the long-range nature of the strain mediated interactions, although more rigorous arguments have been made on the basis of renormalization group theory: using the Ginzburg criterion and the concept of marginal dimensionality $d^*$, Als-Nielsen and Birgeneau[30] described the transition as an example of $d^* = 2$, with the fluctuations in reciprocal space confined to a line. Since the dimensionality $d = 3$, the condition $d > d^*$ leads to the mean-field behavior.[14,30]

For the entire temperature range from 2 to 300 K, PrAlO$_3$ has a larger $C_p$ than LaAlO$_3$.[31] This is attributed to the additional thermal excitations from the Pr$^{3+}$ crystal field levels and the soft acoustic phonons, both playing important roles in the structural phase transitions. Evidently, it is only the singular components of these contributions that make up the $\Delta C_p$ peak, as its entropy of 3.5 J K$^{-1}$mol$^{-1}$ is smaller than the minimum semi-classical value of $R\ln2$ J K$^{-1}$mol$^{-1}$ (where $R$ is the gas constant). Also, previous studies have shown the changes in the Pr$^{3+}$ crystal field levels and in the phonon behavior[7,9] at the structural phase transitions, and this is verified in the $C_p$ curves as a small offset at the $R$–$3c \leftrightarrow Imma$ transition. We note in passing that there is no sign of an anomaly near 118 K, where a strong critical behavior was observed in Brillouin scattering.[32] This is consistent with the results of neutron[15,16] and synchrotron x-ray[15] diffraction studies, which also did not find evidence for a phase transition in this temperature region.

Figure 3 shows the heat capacity divided by temperature, $C_p/T$, for Pr$_{1-x}$La$_x$AlO$_3$ with $x$ = 0, 0.2, 0.4, 0.7, and 0.8. As $x$ increases up to 0.7, the first-order peak at the $R$–$3c \leftrightarrow Imma$ transition and the second-order peak at the $Imma \leftrightarrow C2/m$ transition both become smaller in size. For $x$ = 0.8, only a small anomaly corresponding to the $R$–$3c \leftrightarrow C2/m$ transition is observed at 44 K. A similar trend was reported in the optical absorption measurements,[13] where the splitting of the Pr$^{3+}$ crystal field levels at the transitions diminished with increasing $x$. The transition temperatures also decrease with increasing $x$, and the present results are plotted as filled circles in Fig. 1; our results agree with the highest values reported previously, confirming the high quality of the present crystals. From symmetry consideration, the $R$–$3c \leftrightarrow C2/m$ transition in $x$ = 0.8 is expected to be first order,[18,19] as these space groups are not group-subgroup related.[18,19] However, there is no visible difference between the $C_p$ data taken on heating and cooling directions (Fig. 3, inset), and we found no evidence of latent heat from the relaxation curves. While it is possible that weak first-order features are smeared out at the transition, we note the strong resemblance of the $C_p$ peak to the second-order $Imma \leftrightarrow C2/m$ peak in $x$ = 0.7. Also, the small (but sharp) peak at the $R$–$3c \leftrightarrow Imma$ transition in $x$ = 0.7 became slightly broader for measurements on the cooling direction, so the small size of the $C_p$ peak itself does not explain the absence of first-order characteristics for the $R$–$3c \leftrightarrow C2/m$ transition. It is interesting to notice that the optical absorption data on $x$ = 0.75 show a single transition at ~55 K,[13] but the published result is not detailed enough to determine the order of the transition.

## B. X-ray scattering measurements

To characterize the $R–3c \leftrightarrow C2/m$ transition in more detail, we have tracked the order parameter using high-resolution x-ray scattering; for a first-order transition, some discontinuity in the order parameter can be expected at its onset, and the critical behavior will be different from those expected for second-order transitions. Previously, Birgeneau and co-workers[9] performed neutron scattering measurements on $PrAlO_3$. In addition to providing a detailed mechanism for the second-order $Imma \leftrightarrow C2/m$ transition, these workers showed that the order parameter can be studied accurately by measuring the monoclinic twinning angle $2\delta$ in the $C2/m$ phase.[9] This was experimentally achieved by scanning the crystal orientation angle $\omega$ across the orthorhombic (202) reflection,[9] which splits into (202) and (20–2) in the $C2/m$ structure. Because the orthorhombic (202) originates from the cubic (200),[9,16] which in turn remains a single peak in the $R–3c$ structure, we can study the $R–3c \leftrightarrow C2/m$ transition in $x = 0.8$ by following the splitting pattern of the cubic (200) reflection. The use of a Ge (111) monochromator allows us to carry out x-ray scattering with higher resolution than those typically achieved using neutron scattering.

Reciprocal space maps around the cubic (200) reflection are shown in the left panels of Fig. 4. These maps are plotted as a function of $\omega$ and the scattering angle $2\theta$, such that the diagonal corresponds to a $\theta$-$2\theta$ scan, along the longitudinal direction in reciprocal space. The map at 50 K represents the data in the $R–3c$ phase, where a single peak is observed. On the other hand, the map at 16 K in the $C2/m$ phase shows an additional smaller peak on the lower $\omega$ side. The intensity of the smaller peak is 20% of the main peak at 16 K, while the intensity of the main peak at 16 K is 70% of its value at 50 K. These results provide evidence for majority and minority twin domains in the monoclinic phase, which give rise to Bragg reflections at similar 2θ angle but at different ω.

In order to track the development of monoclinic twinning, we have performed a series of ω scans (with 2θ fixed at 47.85°) in small temperature steps. Representative data are shown in the right panel of Fig. 4. At 45 K in the $R–3c$ phase, only a sharp peak is observed. We find that the peak profile is most accurately reproduced with a Voigt function, which is a convolution between a Gaussian and a Lorentzian function. The result of a least-squares fit to the 45 K data, shown as a solid line, provides a full width at half maximum (FWHM) of 0.04°; with the instrument resolution of less than 0.01°, FWHM in the $R–3c$ phase is determined by the small mosaic spread of the crystal. As the crystal is cooled below the transition, a shoulder appears on the lower ω side, moving farther away with decreasing temperature. At 16 K, the shoulder has become a well-defined peak. When the entire intensity profile is fitted with a single Voigt function, a strong increase in FWHM is found below 44.0 K (Fig. 5, inset). Thus, this is the temperature at which the peak first shows a sign of splitting, coinciding with the peak temperature in $C_p$. Note the sharp onset of increase in FWHM, and any discontinuity, if present, must be below the resolution of the measurements. These results evoke a well-defined second-order transition, and we find no evidence of first-order characteristics from the x-ray scattering measurements.

Having established the onset temperature of peak splitting, we now use two Voigt functions to fit the intensity profile in the $C2/m$ phase. The fit shows excellent agreement with the data below 43.6 K, and the results for the 41, 38, and 16 K data are drawn as solid lines in Fig. 4. One of the parameters obtained from the fit is the splitting angle between the two peaks, which corresponds to the monoclinic twinning angle 2δ.[9] As shown in Fig. 5, 2δ has a typical order-parameter behavior of a second-order transition, continuously approaching 0 as the transition is approached from below. In the lowest temperature region, 2δ approaches a value of 0.13°. This is more than an order of magnitude smaller than 1.57° reported for PrAlO$_3$,[9] reflecting the extremely small monoclinic distortion in $x = 0.8$.

The order parameter is expected to follow a power law behavior $2\delta = 2\delta_0[(T_c - T) / T_c]^\beta$ close to the phase transition,[9] where $T_c$ is the transition temperature and $\beta$ is the critical exponent. To evaluate the values of $T_c$ and $\beta$, log-log plots of $2\delta$ versus reduced temperature are shown in Fig. 6. Also plotted in the figure is the published result of PrAlO$_3$,[9] for which the mean-field critical value ($\beta = 0.5$) is confirmed by the data having the same slope as the solid line. For $x = 0.8$, three sets of plot with different values of $T_c$ are shown, corresponding to a small range of $T_c$'s, ±0.2 K. The results provide the following observations: (1) the three data sets collapse into a constant slope with $\beta = 0.5$ for the reduced temperatures from ~0.06 to ~0.3. (2) Below this region, the plot for $T_c = 43.6$ K shows an upward curvature, indicating an underestimated $T_c$. Similarly, the downward curvature found for $T_c = 44.0$ K indicates that this $T_c$ is overestimated. (3) $T_c = 43.8$ K appears to be the best estimate for $T_c$, showing minimum curvature down to the lowest reduced temperatures.

The regime in Fig. 6 above the reduced temperatures of ~0.06, which is robust to small variations in $T_c$, looks very similar to the data for pure PrAlO$_3$ (albeit with much smaller order parameters). On the other hand, there may be a slight reduction in the slope below the reduced temperatures of ~0.06, perhaps approaching $\beta = 0.367$ of the three-dimensional Heisenberg universality class.[33] Although the increasing scatter and limited data points near $T_c$ do not warrant more quantitative analysis, such a result would be suggestive of a crossover from a mean-field regime to an asymptotic critical regime as the transition temperature is approached. Alternatively, the entire temperature dependence of $2\delta$ is described reasonably well by a full mean-field solution,[34] $m = \tanh(qJm/k_BT)$, with $T_c = 43.8$ K (Fig. 5). The small difference between $T_c = 43.8$ K and the FWHM onset temperature of 44.0 K (Fig. 5, inset) is not unexpected for the solid solution; from the slope of phase boundary in Fig. 1, a concentration gradient of $x = 0.001$ in the crystal is enough to explain a distribution in the transition temperature of 0.2 K. This range of distribution in $T_c$ also accounts for the convergence of the three data sets in Fig. 6 with increasing reduced temperature. In any case, the present result is consistent with the order parameter following the mean-field behavior over a wide temperature region, and in the presence of weak compositional disorder. It is clearly distinguished from weakly first-order transitions where much smaller effective β values (~0.15–0.3; the slope for β = 0.2 is shown in Fig. 6) are often reported,[35–37] a consequence of fitting an order parameter with a weak discontinuity to a continuous function.

We have seen that both heat capacity and x-ray scattering provide no evidence for first-order characteristics at the $R$–$3c$ ↔ $C2/m$ transition. The reason this transition is thought to be first order is that the space group $C2/m$ is not a subgroup of $R$–$3c$,[18,19] with the transition involving an abrupt change in the tilt axis of the AlO$_6$ octahedra. However, these symmetry considerations do not define the size of latent heat, hysteresis, or discontinuity, and it is always possible for the first-order features to be immeasurably small. This certainly appears to be the case for the $R$–$3c$ ↔ $C2/m$ transition in $x$ = 0.8, which is indeed accompanied by a very small monoclinic distortion. The only observation that is not compatible with this argument is the second-order-like critical behavior found in the x-ray scattering measurements. In this regard, it would be of interest to measure the splitting of the Pr$^{3+}$ crystal field levels below $T_c$, not only to check the critical behavior but also to study the coupling of Pr$^{3+}$ electronic degrees of freedom to the structural distortion.

## IV. SUMMARY

We have performed heat capacity and high-resolution x-ray scattering measurements on single crystals of Pr$_{1-x}$La$_x$AlO$_3$. Heat capacity on PrAlO$_3$ shows a sharp first-order peak due to the $R$–$3c$ ↔ $Imma$ transition at 212 K, as well as a classical mean-field peak due to the $Imma$ ↔ $C2/m$ transition at 151 K. Both the heat capacity anomaly and the transition temperature of these transitions diminish with increasing $x$, and only a single $R$–$3c$ ↔ $C2/m$ transition is observed for $x = 0.8$. The lack of group-subgroup relationship between the $R$–$3c$ and $C2/m$ structures should require the transition in $x = 0.8$ to be of first order, but we find no evidence of hysteresis, latent heat, or discontinuity in our heat capacity and x-ray scattering measurements. Our x-ray order parameter measurements on the $x = 0.8$ sample are most consistent with mean-field criticality, with a small distribution of $T_c$ originating from weak compositional disorder. Detailed studies on the temperature dependence of the $Pr^{3+}$ crystal field levels may well clarify the origin and nature of the unusual $R$–$3c$ ↔ $C2/m$ transition in $Pr_{1-x}La_xAlO_3$. We hope that this work will stimulate further efforts to understand the complex structural phase transitions found in perovskite oxides.

## ACKNOWLEDGMENTS

We thank E. Kermarrec for valuable discussion, M. Kiela for technical assistance on x-ray scattering, and S. Takenouchi for performing the composition analysis of the crystals. This study was supported by NSERC of Canada and by a Grant-in-Aid from JSPS under Grand No. 23740245.

# REFERENCES

[1] C. J. Howard, B. J. Kennedy, and B. C. Chakoumakos, J. Phys.: Condens. Matter **12**, 349 (2000).

[2] J. D. Axe, G. Shirane, and K. A. Müller, Phys. Rev. **183**, 820 (1969).

[3] E. Cohen, L. A. Risberg, W. A. Nordland, R. D. Burbank, R. C. Sherwood, and L. G. Van Uitert, Phys. Rev. **186**, 476 (1969).

[4] R. D. Burbank, J. Appl. Cryst. **3**, 112 (1970).

[5] R. T. Harley, W. Hayes, A. M. Perry, and S. R. P. Smith, J. Phys. C: Solid State Phys. **6**, 2382 (1973).

[6] G. A. Gehring and K. A. Gehring, Rep. Prog. Phys. **38**, 1 (1975).

[7] J. K. Kjems, G. Shirane, R. J. Birgeneau, and L. G. Van Uitert, Phys. Rev. Lett. **31**, 1300 (1973).

[8] E. Cohen, M. D. Sturge, R. J. Birgeneau, E. I. Blount, L. G. Van Uitert, and J. K. Kjems, Phys. Rev. Lett. **32**, 232 (1974).

[9] R. J. Birgeneau, J. K. Kjems, G. Shirane, and L. G. Van Uitert, Phys. Rev. B **10**, 2512 (1974).

[10] K. B. Lyons, R. J. Birgeneau, E. I. Blount, and L. G. Van Uitert, Phys. Rev. B **11** 891 (1975).

[11] M. D. Sturge, E. Cohen, L. G. Van Uitert, and R. P. van Stapele, Phys. Rev. B **11**, 4768 (1975).

[12] R. T. Harley, W. Hayes, A. M. Perry, and S. R. P. Smith, J. Phys. C: Solid State Phys. **8**, L123 (1975).

[13] T. J. Glynn, R. T. Harley, W. Hayes, A. J. Rushworth, and S. H. Smith, J. Phys. C: Solid State Phys. **8**, L126 (1975).

[14] R. M. White and T. H. Geballe, *Long Range Order in Solids* (Academic Press, New York, 1979).

[15] S. M. Moussa, B. J. Kennedy, B. A. Hunter, C. J. Howard, and T. Vogt, J. Phys.: Condens. Matter **13**, L203 (2001).

[16] M. A. Carpenter, C. J. Howard, B. J. Kennedy, and K. S. Knight, Phys. Rev. B **72**, 024118 (2005).

[17] M. A. Carpenter, E. C. Wiltshire, C. J. Howard, R. I. Thomson, S. Turczynski, D. A. Pawlak, and T.

Lukasiewicz, Phase Transitions **83**, 703 (2010).

[18] C. J. Howard and H. T. Stokes H T, Acta Cryst. B **54**, 782 (1998); **58**, 565 (2002).

[19] C. J. Howard and H. T. Stokes, Acta Cryst. A **61**, 93 (2005).

[20] M. A. Carpenter and C. J. Howard, Acta Cryst. B **65**, 134 (2009).

[21] M. A. Carpenter and C. J. Howard Acta Cryst. B **65**, 147 (2009).

[22] B. J. Kennedy, C. J. Howard, A. K. Prodjosantoso, and B. C. Chakoumakos, Appl. Phys. A: Mater. Sci. Process. **74**, S1660 (2002).

[23] T. Basyuk, L. Vasylechko, S. Fadeev, I. I. Syvorotka, D. Trots, and R. Niewa, Radiat. Phys. Chem. **78**, S97 (2009).

[24] M. A. Carpenter, R. E. A. McKnight, C. J. Howard, Q. Zhou, B. J. Kennedy, and K. S. Knight, Phys. Rev. B **80**, 214101 (2009).

[25] R. I. Thomson, J. M. Rawson, C. J. Howard, S. Turczynski, D. A. Pawlak, T. Lukasiewicz, and M. A. Carpenter, Phys. Rev. B **82**, 214111 (2010).

[26] M. A. Carpenter, A. Buckley, P. A. Taylor, R. E. A. McKnight, and T. W. Darling, J. Phys.: Condens. Matter **22**, 035406 (2010).

[27] B. M. Wanklyn, S. H. Smith, and G. Garton, J. Crystal Growth **33**, 150 (1976).

[28] V. Hardy, Y. Bréard, and C. Martin, J. Phys.: Condens. Matter **21**, 075403 (2009).

[29] R. Escudero, F. Morales, and S. Bernès, J. Phys.: Condens. Matter **21**, 325701 (2009).

[30] J. Als-Nielsen and R. J. Birgeneau, Am. J. Phys. **45**, 554 (1977).

[31] W. Schnelle, R. Fischer, and E. Gmelin, J. Phys. D: Appl. Phys. **34**, 846 (2001).

[32] P. A. Fleury, P. D. Lazay, and L. G. Van Uitert, Phys. Rev. Lett. **33**, 492 (1974).

[33] M. F. Collins, *Magnetic Critical Scattering* (Oxford University Press, London, 1989).

[34] K. C. Rule, M. J. Lewis, H. A. Dabkowska, D. R. Taylor, and B. D. Gaulin, Phys. Rev. B **77**, 134116

(2008).

[35] U. J. Nicholls and R. A. Cowley, J. Phys.C: Solid State Phys. **20**, 3417 (1987).

[36] U. J. Cox, J. Phys.: Condens. Matter **1**, 3565 (1989).

[37] P. Daniel, M. Rousseau, and J. Toulouse Phys. Rev. B **55**, 6222 (1997).

Figure Captions

**FIG. 1.** Low temperature phase diagram of $Pr_{1-x}La_xAlO_3$. The filled circles correspond to the results of the present study, whereas other symbols are from Glynn et al.,[13] Basyuk et al.,[23] and Carpenter et al.[24]

**FIG. 2.** Heat capacity of $PrAlO_3$ and $LaAlO_3$. The dashed line, obtained by fitting the $C_p$ well above and below the anomalous region by a polynomial function, is the baseline used to determine the excess part of the heat capacity, $\Delta C_p$, shown in the inset.

**FIG. 3.** (Color online) Heat capacity divided by temperature, $C_p/T$, for $Pr_{1-x}La_xAlO_3$ ($x$ = 0, 0.2, 0.4, 0.7, and 0.8) measured on heating direction. The data have been offset by 2.0, 1.5, 1.0, 0.5, and 0 J K$^{-2}$mol$^{-1}$, respectively, for clarity. The inset shows the $x$ = 0.8 data near the $R$–$3c$ ↔ $C2/m$ transition. The filled circles correspond to the data taken on heating direction, while the empty circles correspond to the data taken on cooling.

**FIG. 4.** (Color online) (Left) $2\theta$ versus $\omega$ reciprocal space maps around the cubic (200) reflection at 50 and 16 K for $x$ = 0.8. (Right) The intensity profile of the $\omega$ scans through (200) at selected temperatures. The solid lines are the results of the fit using a single Voigt function (for the data at 45 K) and two Voigt functions (for the data at 41, 38, and 16 K). The intensities for the 38, 41, and 45 K data have been offset for clarity.

**FIG. 5.** Temperature dependence of monoclinic twinning angle $2\delta$ for $x$ = 0.8. The solid line corresponds to the fit to a full mean-field solution. The inset shows the temperature dependence of the full width at half

maximum (FWHM) for the ω scans through the cubic (200) reflection.

**FIG. 6.** (Color online) Log-log plot of 2δ versus reduced temperature for $PrAlO_3$ and $x = 0.8$. The data for $PrAlO_3$ are reproduced from Birgeneau et al.[9] For $x = 0.8$, plots with three different values of $T_c$ are shown. The solid lines correspond to the mean-field critical exponent $\beta = 0.5$, the dashed line corresponds to $\beta = 0.367$ for the three-dimensional Heisenberg universality class, and the dotted line corresponds to $\beta = 0.2$.

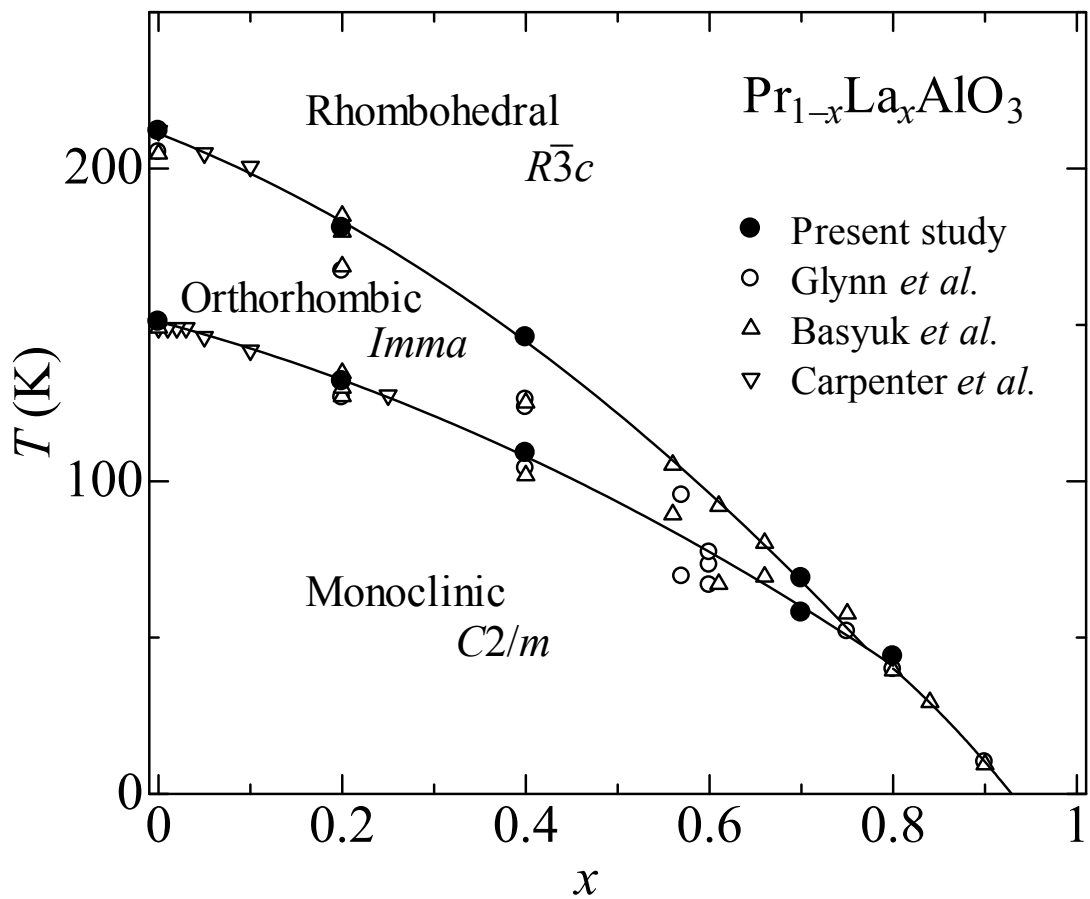

Fig. 1. Tachibana et al.

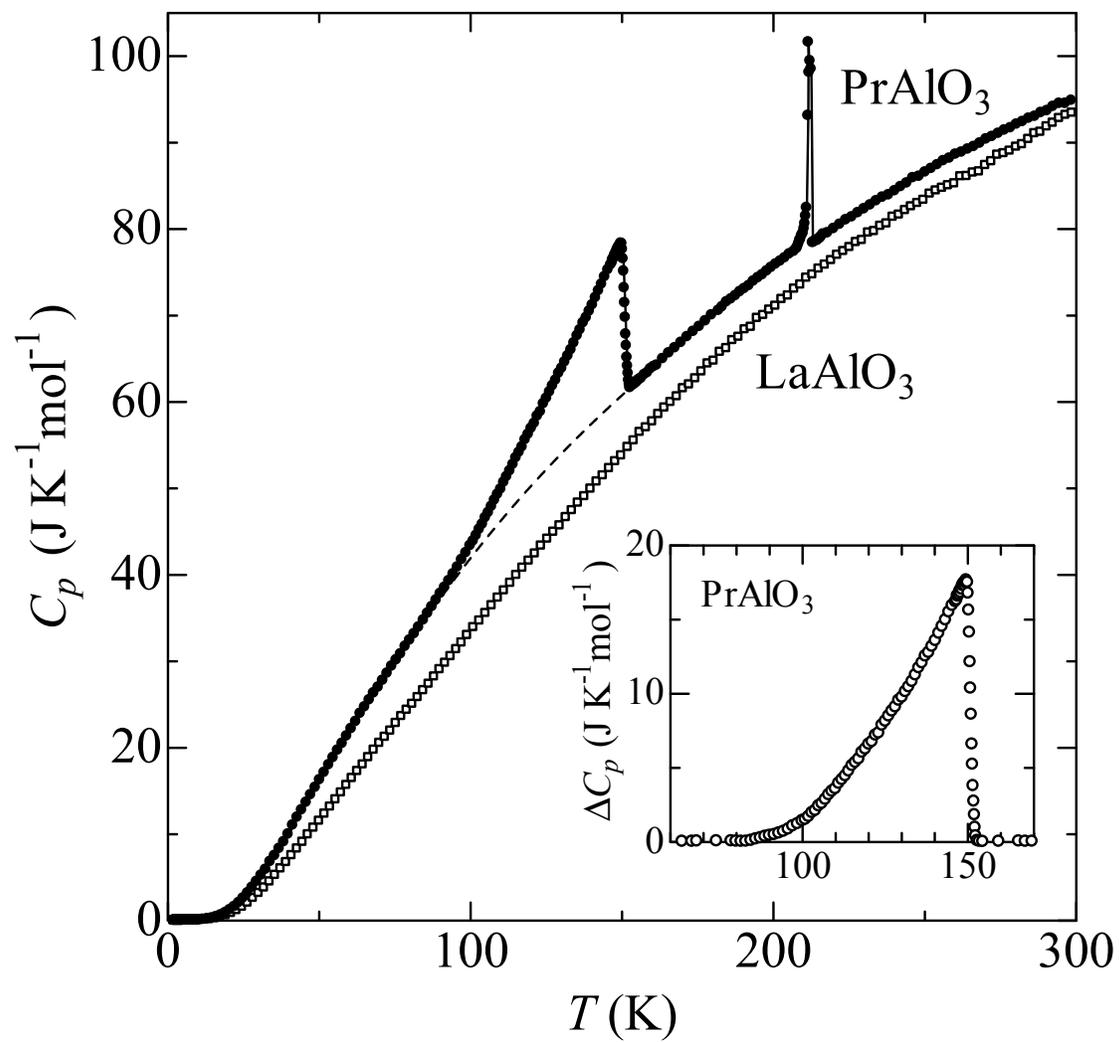

Fig. 2. Tachibana et al.

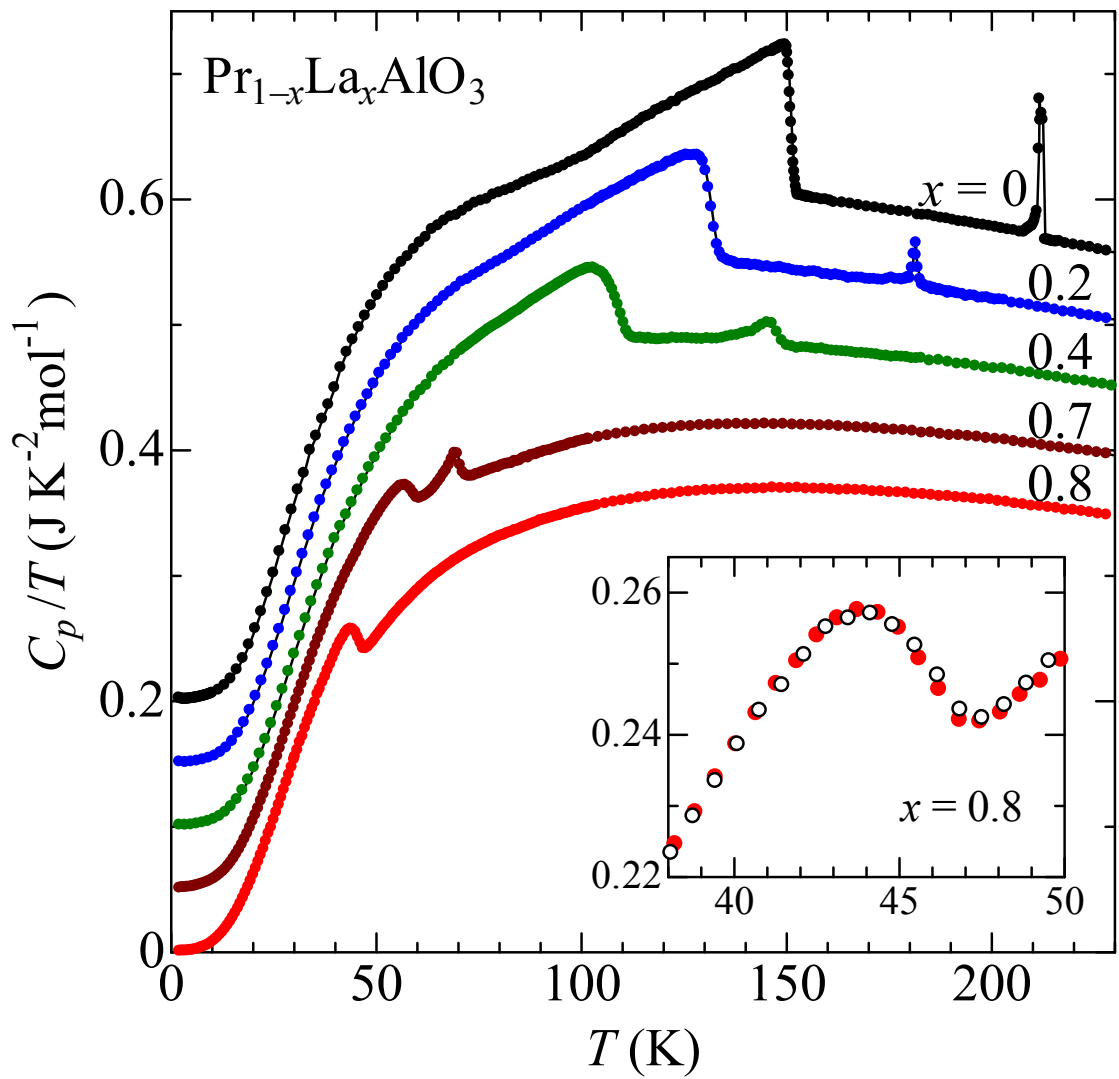

Fig. 3. Tachibana et al.

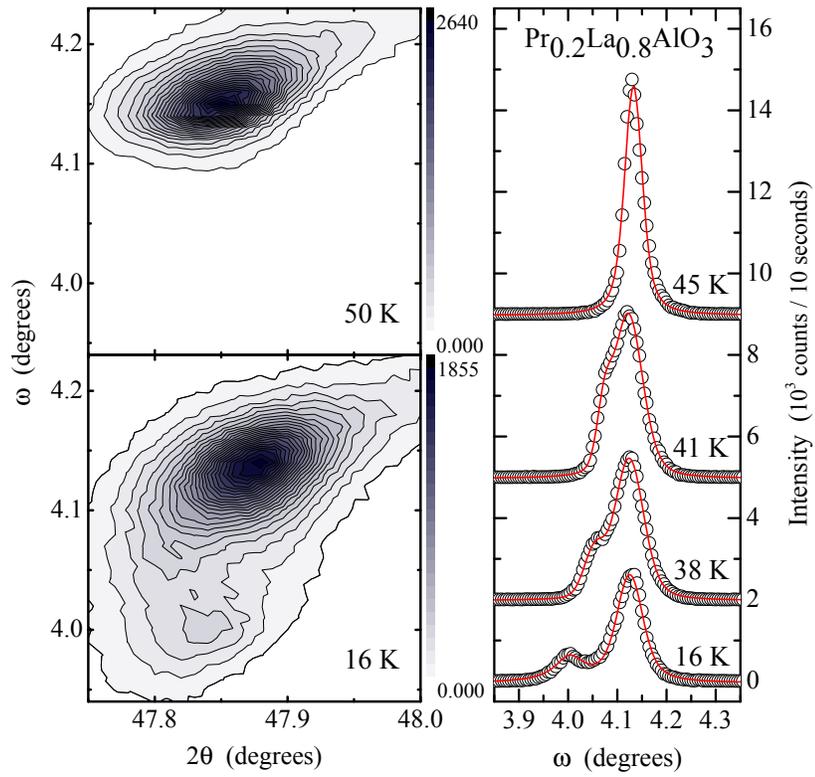

Fig. 4. Tachibana et al

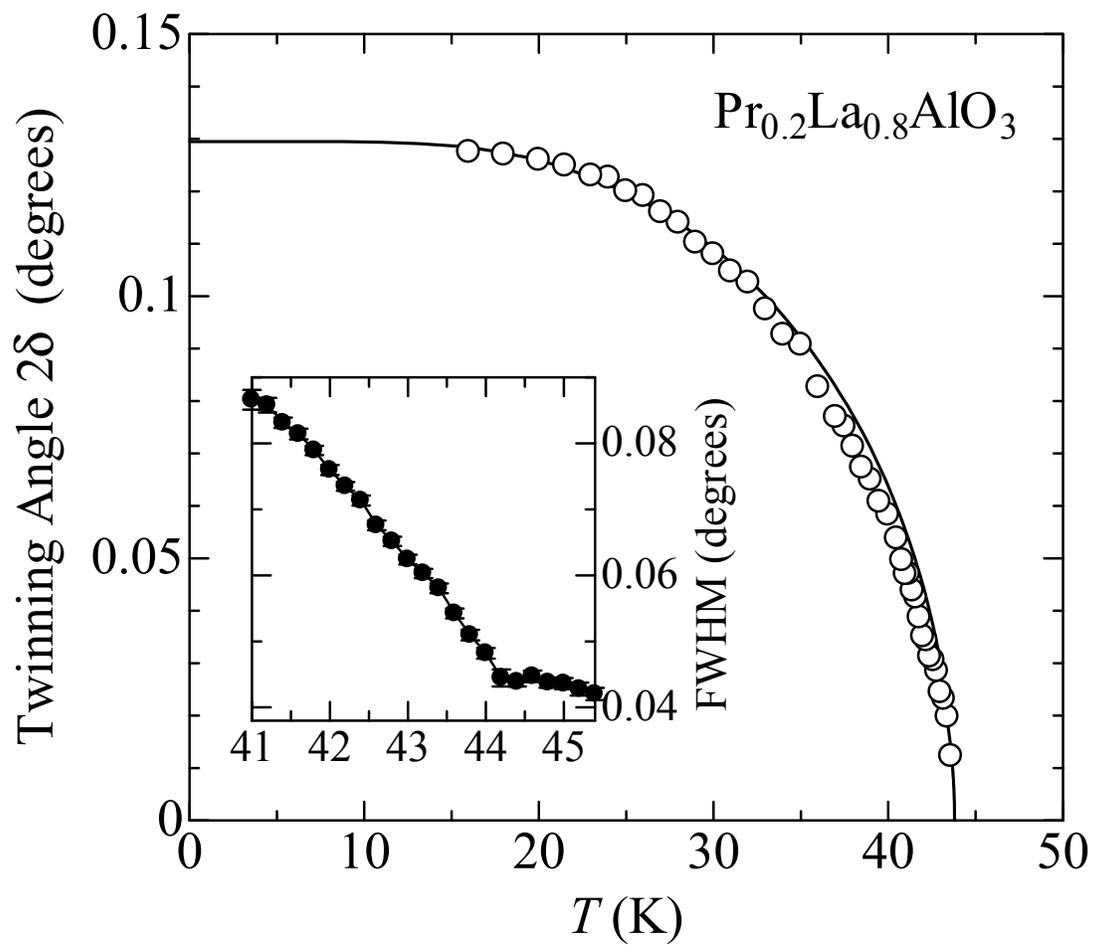

Fig. 5. Tachibana et al.

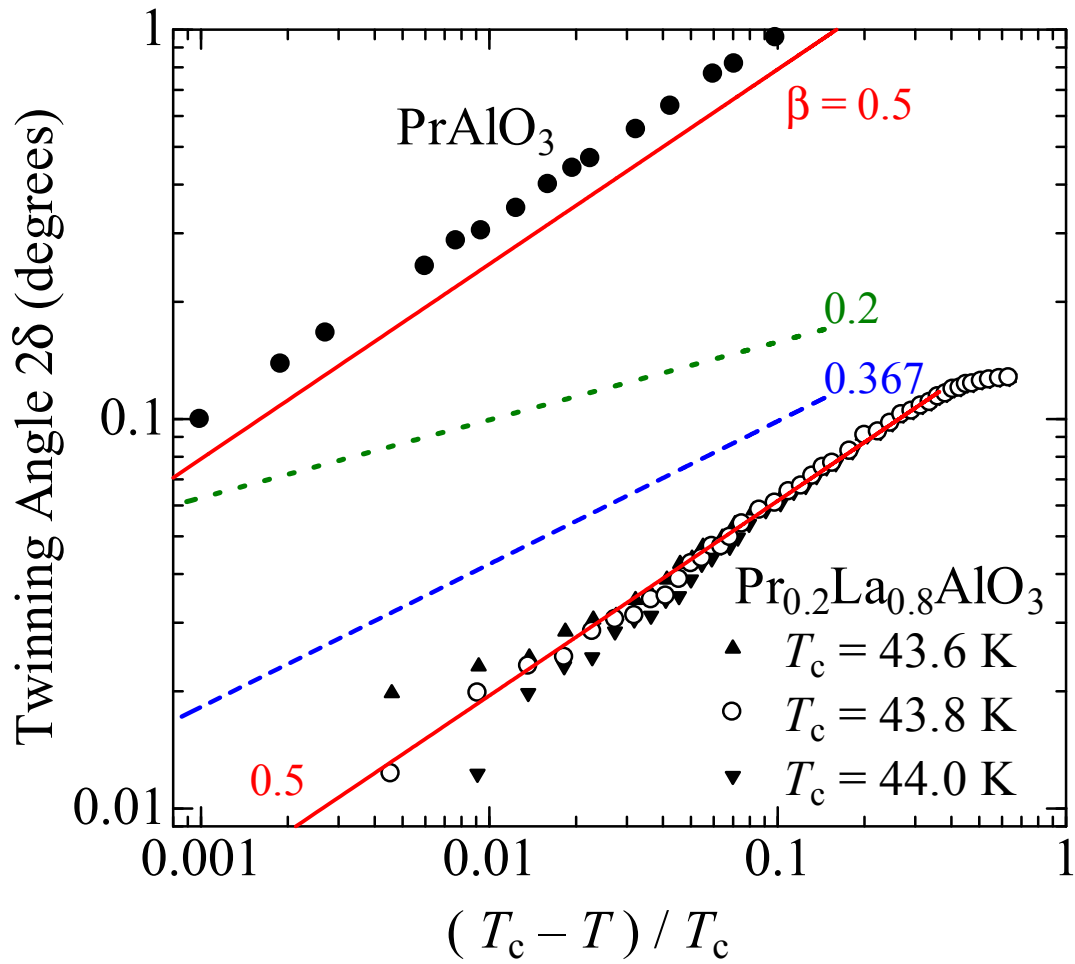

Fig. 6. Tachibana et al.